\documentclass{aa}
\usepackage{graphicx}
\begin{document}
   \title{Multi-wavelength study of the Seyfert 1 galaxy NGC 3783 with XMM-Newton}

   \author{A. J. Blustin
          \inst{1}
          \and
          G. Branduardi-Raymont
          \inst{1}
          \and
          E. Behar
          \inst{2}
          \and
          J. S. Kaastra
	  \inst{3}
          \and
	  S. M. Kahn
	  \inst{2}
	  \and
          M. J. Page
          \inst{1}
          \and
	  M. Sako
	  \inst{4}
	  \and
	  K. C. Steenbrugge
	  \inst{3}
						  }

   \offprints{A. J. Blustin\\
             \email{ajb@mssl.ucl.ac.uk}}

   \institute{MSSL, University College London,
             Holmbury St. Mary, Dorking, Surrey RH5 6NT, England
             \and
             Department of Physics, Columbia University, 550 West 120th Street, New York, NY 10027, USA
             \and
             SRON National Institute for Space Research, Sorbonnelaan 2, 3584 CA Utrecht, Netherlands
             \and
	     Caltech, Pasadena, CA 91125, USA
	  }

   \date{Received 8 May 2002; accepted 14 June 2002}

   \abstract{
We present the analysis of multi-wavelength XMM-Newton data from the
   Seyfert galaxy NGC 3783, including UV imaging, X-ray and UV lightcurves,
   the 0.2$-$10~keV X-ray continuum, the iron K$\alpha$ emission line, and
   high-resolution spectroscopy and modelling of the soft X-ray warm absorber.
    The 0.2$-$10~keV spectral continuum can be well reproduced by a
    power-law at higher energies; we detect a prominent Fe K$\alpha$ emission line, with
    both broad and narrow components, and a weaker emission line at 6.9 keV
      which is probably a combination of Fe K$\beta$ and \ion{Fe}{xxvi}.

We interpret the significant deficit of counts in the soft X-ray region as
 being due to absorption by ionised gas in the line of sight. This is
 demonstrated by the large number of narrow absorption lines in the RGS
 spectrum from iron, oxygen, nitrogen, carbon, neon, argon, magnesium,
 silicon and sulphur. The wide range of iron states present in the spectrum
 enables us to deduce the ionisation structure of the absorbing medium. We
 find that our spectrum contains evidence of absorption by at least two phases of gas: a hotter phase containing plasma with a log ionisation parameter $\xi$ (where $\xi$ is in erg cm s$^{\rm -1}$) of 2.4 and greater, and a cooler phase with log $\xi$ centred around 0.3. The gas in both phases
 is outflowing at speeds of around 800 km s$^{\rm -1}$. The main
 spectral signature of the cold phase is the Unresolved Transition Array
 (UTA) of M-shell iron, which is the deepest yet observed; its depth
 requires either that the abundance of iron, in the cold phase, is several
 times that of oxygen, with respect to solar abundances, or that the absorption lines associated with this phase are highly saturated. The cold phase is associated with ionisation states that would also absorb in the UV.

   }

   \maketitle
%

\section{Introduction}

NGC 3783 is a Seyfert 1 galaxy at redshift 0.00973. Many observers have
found a deficit of counts under its power-law X-ray continuum in the
soft X-ray range, which they interpret as photoelectric absorption by
ionised gas in our line of sight to the active nucleus of the galaxy
(e.g. George et al. \cite{george1998}) - a warm absorber. It has also
been claimed, alternatively (Ghosh et al. \cite{ghosh}), that there is
an excess of counts over the power law in the soft X-ray range. De Rosa et al. (\cite{derosa}), using data from a $\sim$ five day observation with BeppoSAX in 1998, detect both a soft excess and warm absorption in their soft X-ray spectrum. They also found a high-energy cut-off of the power-law continuum at 340 $^{\rm +560}_{\rm -107}$~keV.

With the advent of XMM-Newton and Chandra it has become possible, due
to the high resolution of the X-ray spectrometers carried on these
missions, to study warm absorbers in unprecedented detail, as was evident
after the first high resolution spectrum of a Seyfert galaxy became available
(Kaastra et al. \cite{kaastra2000a}). The first high resolution X-ray spectrum of NGC 3783 (Kaspi et al. \cite{kaspi2000}) was taken in January 2000 using the Chandra
HETGS, with an exposure time of 56 ks, and showed a large number of narrow absorption lines from the H-like and He-like ions of O, Ne, Mg, Si, S and Ar, as well as L-shell transitions of \ion{Fe}{xvii}$-$\ion{Fe}{xxi}, and a few weak emission lines mainly from O and Ne. The blueshifts of the absorption lines indicated that the warm absorber was outflowing at -440 $\pm$ 200 km s$^{\rm -1}$. Further analysis of this dataset
(Kaspi et al. \cite{kaspi2001}) also confirmed the presence of \ion{Fe}{xxii} and \ion{Fe}{xxiii}, and revised the estimate of the absorber's blueshift to -610 $\pm$ 130 km s$^{\rm -1}$; the emission lines were at the systemic velocity of the galaxy. Using regions of the continuum where line absorption is not present, a continuum model was fitted, which required deep \ion{O}{vii} and \ion{O}{viii} absorption edges (implying an N$_{\rm H}$ of order 10$^{\rm 22}$ cm$^{\rm -2}$). The absorption and emission in the gas were then modelled using photoionisation calculations. The model used involved two phases of gas at different levels of ionisation (with an order of magnitude difference between the ionisation parameters) and with different global covering factors. 

NGC 3783 is known
to show UV absorption lines intrinsic to the active nucleus
(Maran et al. \cite{maran}), and there is much interest in attempting
to connect the X-ray and UV warm absorbers in this object (e.g. Shields
\& Hamann \cite{shields}). Kaspi et al. (\cite{kaspi2001}) concluded that the lower-ionisation component in their warm absorber model could give rise to the UV absorption observed in this source, but that this was highly sensitive to the unobservable UV-to-X-ray continuum.

Kaspi et al. (\cite{kaspi2001}) also obtained the first high resolution spectrum of the region around the Fe K$\alpha$ emission line, as the HETGS range extends to these energies. They showed that the Fe K$\alpha$ line was narrow, unresolved even at the HETGS resolution, implying that it originated in the torus region of the AGN. Nandra et al. (\cite{nandra}) had previously fitted a relativistically broadened
Fe K$\alpha$ line to ASCA spectra of NGC 3783, and the De Rosa et al. (\cite{derosa}) analysis of BeppoSAX data detects the presence of both broad and narrow components to the line. 

The most recent Chandra paper on NGC 3783 (Kaspi et al. \cite{kaspi2002}) presents a 900 ks exposure obtained from several observations between February and June 2001, which produced an outstandingly high signal-to-noise HETGS spectrum, the best yet published from an AGN source. H-like and He-like Ca, H-like C, various more lowly ionised states of Si and S, iron species probably up to \ion{Fe}{xxv} and cooler M-shell iron are added to the list of ions detected. The mean blueshift relative to the systemic velocity was found to be -590 $\pm$ 150 km s$^{\rm -1}$, with the absorption lines of many ions being resolved to have a mean FWHM of 820 $\pm$ 280 km s$^{\rm -1}$. The profiles of the absorption lines show asymmetry which, at least in the case of \ion{O}{vii}, originates from the presence of two absorbing systems whose velocity shift and FWHM are consistent with those identified in the UV. Evidence has also been found from this dataset by Behar \& Netzer (\cite{behar2002}), using the inner-shell absorption lines of silicon, that the warm absorber has a continuous distribution of ionisation parameters, and not just two discrete phases. The narrow Fe K$\alpha$ line is now resolved to have a FWHM of 1720 $\pm$ 360 km s$^{\rm -1}$, and it is accompanied by a Compton shoulder redwards of the line itself, though still by no relativistically (or otherwise) broadened component.

This paper describes the results of work on a 40 ks observation
of NGC 3783 with the XMM-Newton Observatory. We obtained UV imaging and photometry from the Optical Monitor (OM; Mason et al. \cite{mason}) simultaneously with X-ray data from the EPIC-PN (Str\"{u}der et al. \cite{struder}) and RGS (den Herder et al. \cite{denherder2001}) instruments. The EPIC cameras, although they only have CCD spectral resolution, have a far larger effective area than HETGS around 6.4 keV where Fe K$\alpha$ emission is observed, requiring far shorter exposures to gain high-statistics spectra at these energies. Again, although the spectral resolution of the RGS is slightly less than that of the HETGS, the combined effective area of the RGS is several times that of the HETGS over much of the energy band where they are both sensitive. Also, the RGS bandpass extends up to about 38~${\rm \AA}$ (0.326~keV), whereas the HETGS effective area falls off quickly past 23~${\rm \AA}$ (0.539~keV), so we can make important additions to the large amount of knowledge already gained on NGC 3783 by Kaspi et al. (\cite{kaspi2000}, \cite{kaspi2001} and \cite{kaspi2002}).

Partly motivated by the lower statistical quality and resolution of our current RGS spectrum, which makes it harder to disentangle the many absorption features displayed in the 900 ks Chandra dataset, we have concentrated on self-consistent global fitting of the warm absorber rather than detailed measurements of the individual absorption lines. The RGS spectrum of NGC 3783 stands on the threshold between traditional X-ray astronomy - where spectroscopy relied on the $\chi$$^{\rm 2}$ fitting of models to often low spectral resolution and poor statistics data - and traditional optical astronomical spectroscopy, where the instrumental resolution and collecting area are sufficiently good to allow precision measurements of individual spectral features. Analysing an RGS spectrum as complex as this requires far more careful usage of $\chi$$^{\rm 2}$ fitting techniques than with spectra of lower resolution; there are so many datapoints, and so many interlinked parameters in the increasingly complex spectral models which are fitted to them, that the practical considerations in the global modelling of such spectra will become increasingly important to X-ray astronomers who work on high-resolution AGN spectroscopy. In this paper, within our analysis of the RGS spectrum of NGC 3783, we attempt to define a practical methodology for the modelling and fitting of such datasets.

This work is a prelude to the analysis of a 280 ks XMM-Newton observation of NGC 3783, which has now been carried out, that will provide an RGS spectrum with equivalent signal-to-noise to the 900 ks HETGS spectrum. This new dataset will extend to higher wavelengths the range which has already been observed by the HETGS, and, since it will be accompanied by high time resolution UV photometry by the OM, will be used to investigate further the multiwavelength variability of this AGN. The long observation will also enable time-resolved spectroscopy of the Fe K$\alpha$ line region.


\section{Observations and data analysis}

NGC 3783 was observed by XMM-Newton, in an RGS Guaranteed Time observation,
on 28$-$29$^{th}$ December 2000 for a total of about 40 ks.
The instrument modes and total exposure times are given in Table~\ref{modes}.

   \begin{table*}
    
      \caption[]{Instrument modes and exposure times.}
         \label{modes}
     $$
         \begin{array}{p{1.5in}p{2in}p{2in}r}
            \hline
            \noalign{\smallskip}
            Instrument      &  Mode & Exposure time (s) \\
            \noalign{\smallskip}
            \hline
            \noalign{\smallskip}
            EPIC-MOS 1 & Large Window (medium filter) & 37724    \\
            EPIC-MOS 2 & Large Window (medium filter) & 37724       \\
            EPIC-PN & Small Window (medium filter) & 37263  \\
            RGS 1 & Spectroscopy & 40456             \\
            RGS 2 & Spectroscopy & 40456             \\
            OM & Image (UVW2 filter) & 40016            \\
            \noalign{\smallskip}
            \hline
         \end{array}
          $$   
  \end{table*}

The MOS1 and MOS2 datasets were processed using \verb/emchain/ under SAS
Version 5.2. Both were found to be piled-up, so the X-ray lightcurve,
spectral continuum, and parameters of the Fe K$\alpha$
line were obtained from the PN data, processed using \verb/epproc/ under SAS
Version 5.2. Since the PN was operated in Small Window Mode, the observed count rate
 ($\sim$ 19 counts s$^{\rm -1}$ before background subtraction) was well below the
 threshold (130 counts s$^{\rm -1}$, XMM-Newton User's Handbook (\cite{uhb})) where pile-up
 becomes a problem. The PN spectrum and lightcurve were extracted from within
a circular region 1\arcmin \, in radius, centred on the source. The background spectrum and
lightcurve were extracted from a region of the same size 3.1\arcmin \, away. The background-subtracted count
rate of the Seyfert nucleus was 18.73 $\pm$ 0.03 counts s$^{\rm -1}$ over the PN band of 0.1$-$12~keV. The spectrum
was analysed in XSPEC and, since the data were taken in small window mode,
a response matrix optimised for this mode, epn\_sw20\_sY9\_medium.rmf
(provided by F. Haberl) was used. The lightcurve was analysed using the XRONOS suite of tools.

The RGS1 and RGS2 spectra were extracted using \verb/rgsproc/ under SAS V5.2.
Background-subtraction is performed with the SAS using regions adjacent to those containing the source in the spatial and spectral domains. The two spectra
were then combined; the RGS1 and RGS2
response matrices are corrected for the differences in effective area
between the two instruments
using the residuals of a power law fit to a pure continuum source, the BL Lac
Mrk 421. This also corrects for the instrumental absorption edge due to
neutral oxygen. The two response matrices - and spectra - can then be
added channel by channel. The resulting spectrum was analysed using
SPEX 2.00 (Kaastra et al. \cite{kaastra2002a}).

\begin{figure}
   \centering
   \includegraphics[width=8cm]{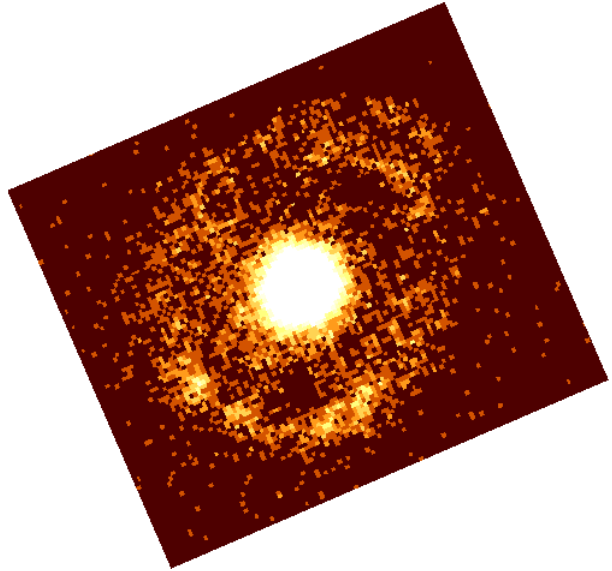}
      \caption{UV image of NGC 3783; North is up, East to the left, and the image is 
               about 50 x 40\arcsec \, in size. The nuclear image is saturated in order to show the spiral arms.
              }
         \label{uv_image}
   \end{figure}

The OM was operated in Rudi-5 imaging mode during our 
observation. This gave a series of 30 UVW2 filter (140$-$270~nm) images of
the galaxy at 0.5\arcsec \, spatial resolution, which were combined using FTOOLS
to form the image in Fig.~\ref{uv_image}. This is the first image
to be taken of this galaxy in this waveband and shows much UV emission from the spiral
arms, about 30\arcsec \, in
diameter, as well as from the nucleus. The FWHM of the central source is about 2.5\arcsec \, which is consistent with the OM Point Spread Function (PSF) with the UVW2 filter.

The long series of images also allowed us to produce a UV lightcurve for the AGN. They were processed using the \verb/omichain/ routine under a development version of SAS (xmmsas\_20020107\_1901-no-aka), and photometry of the active nucleus was performed with GAIA using an aperture of 10\arcsec \, diameter. Background subtraction was performed using counts extracted from an annular region around the source, avoiding the spiral arms, and deadtime and coincidence loss corrections were applied to the count rates. The resulting lightcurve was analysed using the XRONOS tools.

%


\section{Lightcurves}

\begin{figure}
   \centering
   \includegraphics[width=9cm]{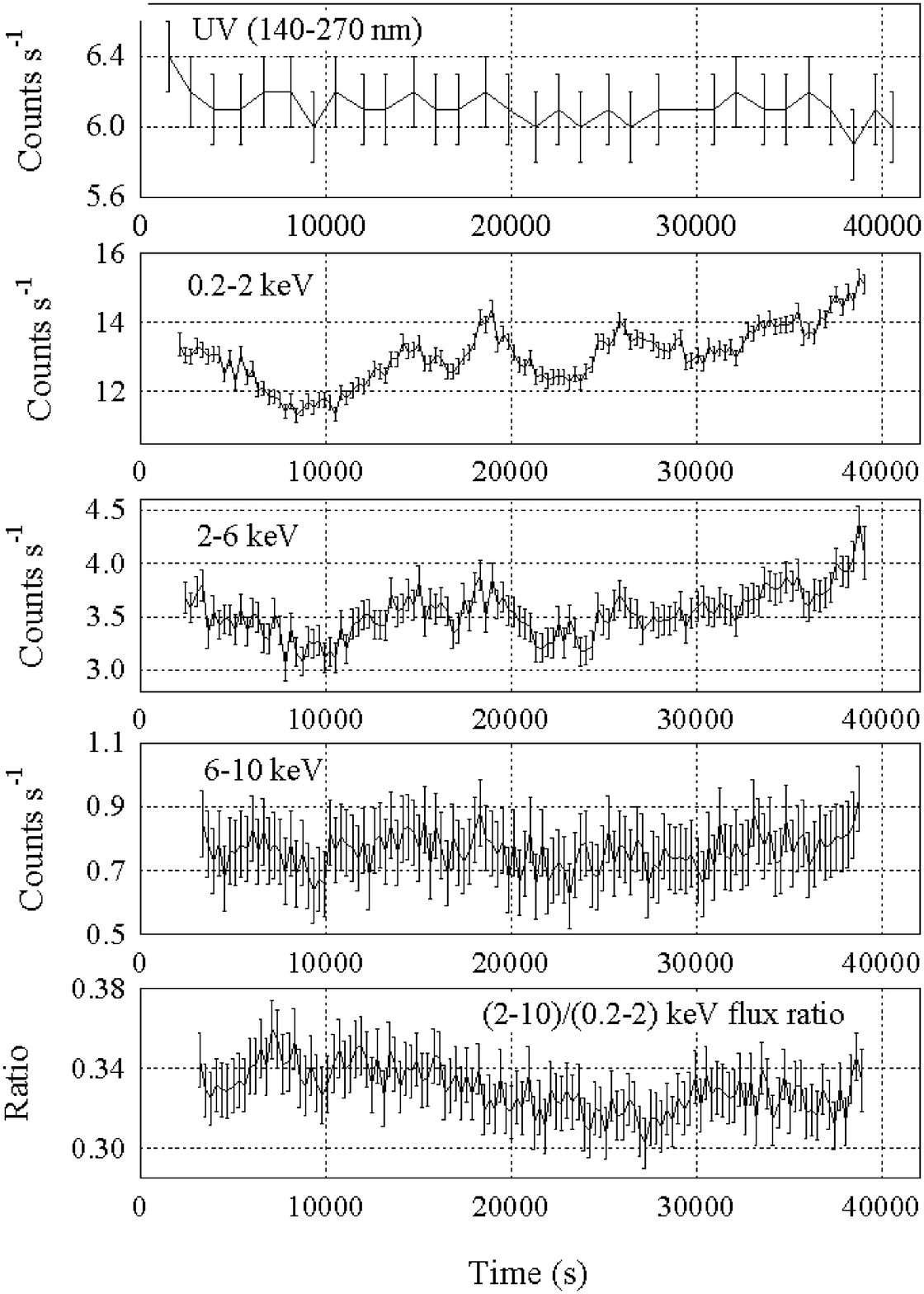}
      \caption{Multiwavelength background-subtracted lightcurves: top, UV (140$-$270~nm) lightcurve from the OM, in 1000 s time bins; middle three, PN lightcurves in the bands 0.2$-$2, 2$-$6 and 6$-$10~keV, in 300 s time bins; bottom, the hardness ratio (2$-$10~keV / 0.2$-$2~keV flux) in 300 s time bins.
              }
         \label{ltcv}
   \end{figure}

The background-subtracted lightcurves of NGC 3783 in different energy bands are shown in Fig.~\ref{ltcv}. The total 0.2$-$10~keV X-ray flux varied by about 30\% between its
highest and lowest points during the course of the observation, whilst the UV flux shows very little variation.

%

   \begin{figure}
   \centering
   \includegraphics[angle=-90,width=8cm]{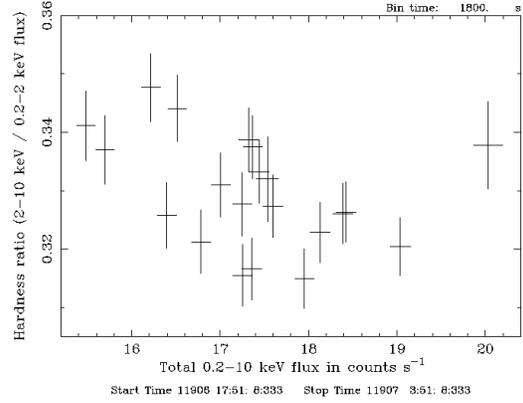}
      \caption{Hardness -- intensity plot in 30 minute time bins.
              }
         \label{colour}
   \end{figure}
%

The variations in the soft (0.2$-$2~keV) and hard (2$-$10~keV) bands follow each other very closely. A plot of the hardness ratio against the total 0.2$-$10~keV intensity (Fig.~\ref{colour}), for 30 minute time bins, may indicate that the PN spectrum gets softer as the source intensity increases. The linear correlation coefficient for the data in this plot is 0.35, with a probability of about 3\% of obtaining this value or greater from a random sample, implying a weak correlation significant at 2 sigma. The final point, at the highest intensity, does not fit in with this trend and corresponds to a sudden hardening of the spectrum at the end of our observation. If this point is disregarded, the softness-intensity correlation is significant at 3 sigma.


\section{Continuum and Fe K$\alpha$ line}

\begin{figure}
   \centering
   \includegraphics[angle=-90, width=8cm]{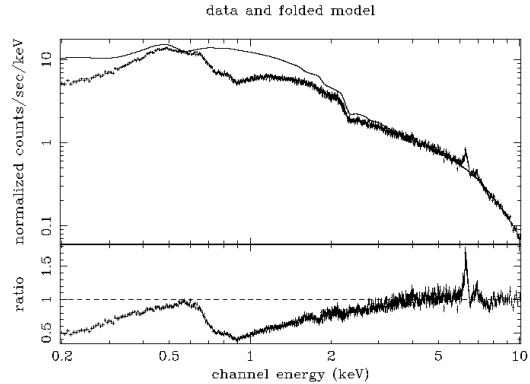}
      \caption{Continuum power-law fit to the 3.5$-$5.5 and 7.5$-$10~keV
      ranges of the PN spectrum, extrapolated over 0.2$-$10~keV. The ratio of the data to the continuum model is shown in the bottom panel.
              }
         \label{cont}
   \end{figure}

To derive the X-ray spectral continuum of NGC 3783, the PN spectrum was
fitted in XSPEC with a power-law model and neutral absorption (using the
tbabs module), with the spectrum binned to 300 counts per bin.
It was found that a simple power-law could not fit the data over 
the whole energy range, due to the significant deficit of counts below
about 3 keV. However, it was possible to produce
an acceptable power-law fit at higher energies, and in the ranges 
3.5$-$5.5 and 7.5$-$10~keV (chosen to avoid the prominent Fe K emission 
around 6.4~keV), the best-fit parameters given in Table~\ref{powerlaw}
were obtained. The neutral absorption was
held at the Galactic value of 8.7 x 10$^{\rm 20}$ cm$^{\rm -2}$ (as in Kaspi et al. \cite{kaspi2001}).
This fit is shown, extrapolated over 0.2$-$10~keV, in Fig.~\ref{cont}, where the form and extent of the warm absorption is clearly visible (although there are still some calibration uncertainties in the PN small window mode, especially below about 0.7~keV). No soft excess is required.

%

With this model, the intrinsic 2$-$10~keV source flux is
8.5 x 10$^{\rm -11}$ erg cm$^{\rm -2}$ s$^{\rm -1}$.

  \begin{table}
    
      \caption[]{The best fit power-law continuum parameters, fitted to the ranges 3.5$-$5.5 and 7.5$-$10 keV.}
         \label{powerlaw}
     $$
         \begin{array}{p{1in}p{1in}p{0.5in}p{0.5in}}
            \hline
            \noalign{\smallskip}
            Photon index ($\Gamma$) &  Normalisation$^{\mathrm{a}}$ & N$_{\rm {H Gal}}$$^{\mathrm{b}}$ & $\chi$$^{\rm 2}_{\rm reduced}$ \\
            \noalign{\smallskip}
            \hline
            \noalign{\smallskip}
            1.60 $\pm$ 0.02 & 0.0180 $\pm$ 0.0007 & 8.7  & 1.4$^{\mathrm{c}}$ \\
            \noalign{\smallskip}
            \hline
         \end{array}
     $$   
\begin{list}{}{}
\item[$^{\mathrm{a}}$] in photons keV$^{\rm -1}$ cm$^{\rm -2}$ s$^{\rm -1}$ at 1 keV
\item[$^{\mathrm{b}}$] in 10$^{\rm 20}$ cm$^{\rm -2}$, fixed
\item[$^{\mathrm{c}}$] for 168 degrees of freedom
\end{list}
  \end{table}

\begin{figure}
   \centering
   \includegraphics[width=9cm]{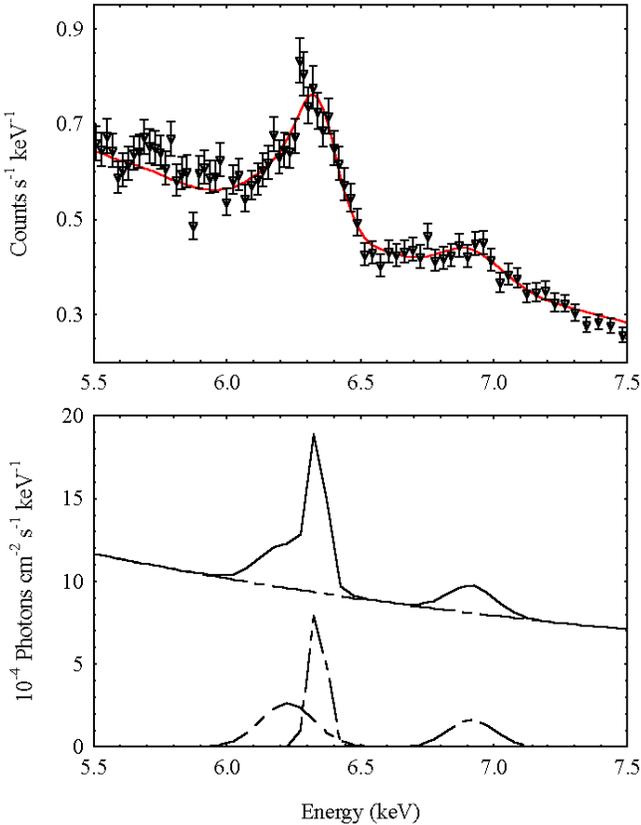}
      \caption{Fit to the iron emission lines in the PN spectrum
      (left, lines 1 and 2 (Fe K$\alpha$); right, line 3). Top panel: data (black) and model (red). Bottom panel: overall model (solid line) and model components (dotted line).
              }
         \label{lines}
   \end{figure}

There is a prominent Fe K$\alpha$ emission line visible in our PN 
spectrum (see Fig.~\ref{cont}). The form of this line implies that it has both broad (line 1) and narrow (line 2) components, and we have fitted each of these as a gaussian, using the power-law of Table~\ref{powerlaw} as the underlying continuum. There is a second, much fainter emission line at a
slightly higher energy (line 3), also parameterised as a gaussian, which could be identified as Fe K$\beta$ or \ion{Fe}{xxvi}. The parameters of the resulting fit are given in Table~\ref{3_lines}; over the range of 5.5$-$7.5~keV, this fit has a $\chi$$^{\rm 2}_{\rm red}$ of 1.1 for 70 degrees of freedom, and is shown in Fig.~\ref{lines}.

  \begin{table*}
    
      \caption[]{The parameters of three gaussians fitted to the emission lines seen in the PN spectrum.}
         \label{3_lines}
     $$
         \begin{array}{p{0.3in}p{1in}p{0.4in}p{0.8in}p{0.6in}p{0.8in}p{0.9in}p{0.8in}}
            \hline
            \noalign{\smallskip}
            Line & Identification & E$_{\rm rest}$$^{\mathrm{a}}$ & E$_{\rm meas}$$^{\mathrm{b}}$ & Flux$^{\mathrm{c}}$ & FWHM$^{\mathrm{d}}$ & V$_{\rm broad}$$^{\mathrm{e}}$ & cz$^{\mathrm{f}}$  \\
            \noalign{\smallskip}
            \hline
            \noalign{\smallskip}
            1 & broad Fe K$\alpha$ & 6.39 & 6.29 $\pm$ 0.03 & 7 $\pm$ 1 & 240 $\pm$ 80 & 11000 $\pm$ 4000 & 5000 $\pm$ 1000 \\
            2 & narrow Fe K$\alpha$ & 6.39 & 6.40 $\pm$ 0.01 & 7.1 $\pm$ 0.9 & 70 $\pm$ 50 & 3000 $\pm$ 3000 & -600 $\pm$ 600 \\
            3 & Fe K$\beta$ & 7.06 & 6.98 $\pm$ 0.04 & 3.7 $\pm$ 0.9 & 200 $\pm$ 100 & 9000 $\pm$ 6000 & 3000 $\pm$ 2000 \\
              & \ion{Fe}{xxvi} & 6.96 &   &   &   &   & -1000 $\pm$ 2000$^{\mathrm{g}}$ \\
            \noalign{\smallskip}
            \hline
         \end{array}
     $$   
\begin{list}{}{}
\item[$^{\mathrm{a}}$] expected rest frame energy in keV
\item[$^{\mathrm{b}}$] measured rest frame energy in keV
\item[$^{\mathrm{c}}$] in 10$^{\rm -5}$ photons cm$^{\rm -2}$ s$^{\rm -1}$
\item[$^{\mathrm{d}}$] Full Width Half Maximum in eV
\item[$^{\mathrm{e}}$] velocity broadening in km s$^{\rm -1}$
\item[$^{\mathrm{f}}$] redshifted velocity in km s$^{\rm -1}$ (negative values indicate blueshifts)
\item[$^{\mathrm{g}}$] if line 3 is identified with \ion{Fe}{xxvi} 
\end{list}
  \end{table*}

The FWHM of the narrow component of our Fe K$\alpha$ line, 3000 $\pm$ 3000 km s$^{\rm -1}$, is consistent with that of Kaspi et al. (\cite{kaspi2002}) (1860 $\pm$ 340 km s$^{\rm -1}$), although it is poorly constrained. The flux in our line is slightly greater, (7.1 $\pm$ 0.9) x 10$^{\rm -5}$ photons cm$^{\rm -2}$ s$^{\rm -1}$ as opposed to (5.26 $\pm$ 0.63) x 10$^{\rm -5}$ photons cm$^{\rm -2}$ s$^{\rm -1}$. The narrow Fe K$\alpha$ line of Kaspi et al. (\cite{kaspi2002}) is not blueshifted; in our case, the blueshift is consistent with zero (-600 $\pm$ 600 km s$^{\rm -1}$), although again this is not well constrained. These authors detect a red wing to their Fe K$\alpha$ line which they identify as a Compton reflection hump, whereas our Fe K$\alpha$ line has a very strong broad component containing as much flux as the narrow component itself. De Rosa et al. (\cite{derosa}) also detect broad and narrow components in the Fe K$\alpha$ line. Their broad component, with $\sigma$ = 0.72 $^{\rm +1.28}_{\rm -0.27}$~keV, is significantly broader than ours ($\sigma$ = 0.10 $\pm$ 0.03~keV). It is also clear, from Fig.~\ref{lines}, that there is an excess of counts over the continuum redwards of Fe K$\alpha$ in our spectrum, and this could correspond to the Compton shoulder observed by Kaspi et al. (\cite{kaspi2002}).

If the broad and narrow components of the Fe K$\alpha$ line are each associated with an Fe K$\beta$ line at the expected $\sim$ 14\% of the flux of K$\alpha$, then the combined flux in K$\beta$ is not sufficient to explain the measured flux in line 3. About half the flux from this line probably originates, then, from \ion{Fe}{xxvi}.

Noting the significant redshift of the broad component of Fe K$\alpha$, it is possible to fit the entire line (or just the broad component) with relativistically broadened emission from an accretion disc. The broad component does not contrast sufficiently against the continuum for the shape of such a line to be determined unambiguously, although it can be fitted with the (almost gaussian) profile of a line from a face-on disc with an emissivity index below about 2.

%


\section{RGS spectrum}

\subsection{Continuum}

The continuum underlying the RGS spectrum was modelled with the power-law fitted to the PN data. The slope and normalisation of this power-law were corrected for the systematic calibration differences between the PN and RGS as described in den Herder et al. (\cite{denherder2002}). The photon index $\Gamma$ became 1.54 and the new normalisation is 0.0175 photons keV$^{\rm -1}$ cm$^{\rm -2}$ s$^{\rm -1}$ at 1 keV, with the Galactic absorption again held constant at 8.7 x 10$^{\rm 20}$ cm$^{\rm -2}$.

With the chosen continuum, there was a significant deficit of flux
below 17~${\rm \AA}$ (above 0.729~keV; see Fig.~\ref{rgs_spec}). There is no excess over the power law,
and the overall form of the spectrum does not contain the extreme
'saw-tooth'
features seen in Mrk 766 and MCG-6-30-15 (Branduardi-Raymont et al.
\cite{branduardi}); however, there is a deep drop in the continuum at
about 17~${\rm \AA}$ which has in
the past been explained as an O VII absorption edge (e.g. George et al.
\cite{george1998}). There is also a
small triangular feature centred at 15~${\rm \AA}$ (0.827~keV). Our task is then to work
out how much of this spectral structure is due to emission, how much to
absorption, and what physical model might explain the spectrum that we
observe.

\subsection{Modelling}

We aim, in modelling this spectrum, to take into account both line and continuum absorption self-consistently. The narrow absorption lines from the photoionised absorber are our starting point; line absorption is a far more reliable indication of the physical state of the absorber than continuum absorption, since the contribution of the many different ions to the continuum absorption is very hard to determine (unless the columns are high enough to produce edges), whereas the lines, if they can be identified, are unambiguous. 

This fact is at the heart of the difficulty with using standard $\chi$$^{\rm 2}$ fitting techniques to model such spectra. Whilst the absorption lines are the best diagnostic of the warm absorption, they are represented by very few datapoints. Continuum absorption, however, is spread out over every datapoint in the spectrum, and this will bias a $\chi$$^{\rm 2}$ minimisation process towards explaining the highly ambiguous continuum absorption rather than the lines themselves, thus providing misleading - and possibly even meaningless - results. This is why $\chi$$^{\rm 2}$ fitting has to be used with great care with spectra such as these, and its use must always accompany a thorough understanding of the physical properties of the model that is being fitted. 

We used SPEX (Kaastra et al. \cite{kaastra2002a}) for the modelling, which provides two possibilities for a photoionised absorber. The \emph{slab} model applies line and continuum absorption by individual ions to a spectral continuum, whilst \emph{xabs} is a grid model of Xstar runs applying absorption by a column of photoionised gas at a range of different ionisation parameters, column densities, and elemental abundances. We define the ionisation parameter $\xi$ here as

   \begin{equation}
      \xi = \frac{L}{n r^{\mathrm{2}}} \,,
   \end{equation}
 
 where $L$ is the source luminosity (in erg s$^{\rm -1}$), $n$ the gas density
 (in cm$^{\rm -3}$) and $r$ the source distance in cm, so $\xi$ has the units
 erg cm s$^{\rm -1}$ (Tarter et al. \cite{tarter}) which are used throughout this paper.
 \emph{Slab} and \emph{xabs} incorporate new
 calculations, using HULLAC (Bar-Shalom et al. \cite{bar}), for
 absorption due to states of M-shell
 iron and lower ionisation states of oxygen (\ion{O}{ii}$-$\ion{O}{vi}). \emph{Slab} is the best tool for detailed measurements of individual ion columns in a spectrum, whereas \emph{xabs} enables the physical parameters and composition of a warm absorber to be investigated directly. A more detailed description of these models is given in Kaastra et al. (\cite{kaastra2002a}).

Our modelling strategy involves the complementary use of these two models. Firstly, \emph{slab} was used to identify the narrow absorption lines, by matching their depths and positions with those corresponding to each ionisation state of each element that the model contains. \emph{Slab} applies both narrow line and continuum absorption self-consistently to the spectrum, so it was possible to see, by looking at each separate ion in turn, the precise effect that the different ions had on the form of the spectral continuum. 

For purposes of comparison with the published spectrum of Kaspi et al. (\cite{kaspi2002}), and with future analyses of RGS spectra, Table~\ref{abs_lines} lists measured and rest wavelengths, blueshifts and equivalent widths for the deepest absorption lines of the most important ions of O, N, C and Fe, as well as these parameters plus the flux of the forbidden emission line of \ion{O}{vii}. It also quotes values of flux and equivalent width measured by Kaspi et al. (\cite{kaspi2002}) where applicable. Note that the deepest line of a particular ion is not always the one theoretically predicted; this is almost certainly because many of the deepest absorption lines are accompanied by re-emission (see Fig.~\ref{spec_1} and Fig.~\ref{spec_2} for an expanded view of tha RGS spectrum and the warm absorber) as demonstrated first by Kaspi et al. (\cite{kaspi2002}). Also, in the case of \ion{O}{vi}, the theoretically predicted deepest line lies in the region of the \ion{O}{vii} emission triplet, making it difficult to distinguish, so our table lists measurements for the second deepest line of \ion{O}{vi}. \ion{C}{vi} Ly-$\beta$ is listed as it occurs in a range where Galactic absorption has less of an effect than at the corresponding Ly-$\alpha$ transition, making it easier to obtain an accurate measurement.  

  \begin{table*}
    
      \caption[]{The blueshifts and equivalent widths of some important spectral lines in the RGS data.}
         \label{abs_lines}
     $$
         \begin{array}{p{0.8in}p{0.9in}p{0.6in}p{0.9in}p{0.8in}p{0.8in}p{0.7in}p{0.9in}}
            \hline
            \noalign{\smallskip}
            Line & $\lambda$$_{\rm {meas}}^{\mathrm{a}}$ & $\lambda$$_{\rm {rest}}^{\mathrm{b}}$ & v$_{\rm blue}$$^{\mathrm{c}}$ & Flux$^{\mathrm{d}}$ & Flux$_{\rm chan}^{\mathrm{e}}$ & EW$^{\mathrm{f}}$ & EW$_{\rm chan}^{\mathrm{g}}$ \\
            \noalign{\smallskip}
            \hline
            \noalign{\smallskip}
\ion{O}{iv} K$\beta$ & 22.67 $\pm$ 0.02 & 22.777$^{\mathrm{h}}$ & -1400 $\pm$ 300 &   &   & 80 $\pm$ 10 & -- \\
\ion{O}{v} K$\alpha$ & 22.27 $\pm$ 0.02 & 22.334 & -900 $\pm$ 300 &   &   & 70 $\pm$ 20 & -- \\
\ion{O}{vi} K$\alpha$$^{\mathrm{i}}$ & 19.293 $\pm$ 0.006 & 19.341 & -760 $\pm$ 90 &   &   & 76 $\pm$ 9 & -- \\ 
\ion{O}{vii} K$\alpha$$^{\mathrm{j}}$ & 21.5 $\pm$ 0.1 & 21.602 & -900 $\pm$ 1000 &   &   & 20 $\pm$ 20 & 40.1 $\pm$ 34.6 \\ 
\ion{O}{viii} Ly-$\alpha$$^{\mathrm{k}}$ & 18.92 $\pm$ 0.01 & 18.969 & -900 $\pm$ 200 &   &   & 50 $\pm$ 10 & 53.6 $\pm$ 15.9 \\ 
\ion{N}{vii} Ly-$\alpha$ & 24.67 $\pm$ 0.06 & 24.781 & -1300 $\pm$ 700 &   &   & 90 $\pm$ 10 & 72.6 $\pm$ 34.9 \\ 
\ion{C}{vi} Ly-$\beta$$^{\mathrm{l}}$ & 28.39 $\pm$ 0.01 & 28.466 & -800 $\pm$ 200 &   &   & 100 $\pm$ 10 & -- \\ 
\ion{O}{vii} (f) & 22.06 $\pm$ 0.02 & 22.101 & -600 $\pm$ 200 & 7 $\pm$ 2 & 10.13 $\pm$ 2.74  & -130 $\pm$ 40 & -270.4 $\pm$ 73.2 \\ 
\ion{Fe}{xvii} & 14.97 $\pm$ 0.02 & 15.014 & -800 $\pm$ 400 &   &   & 92 $\pm$ 6 & 26.2 $\pm$ 5.3 \\ 
\ion{Fe}{xviii} & 14.17 $\pm$ 0.01 & 14.210$^{\mathrm{m}}$ & -800 $\pm$ 200 &   &   & 148 $\pm$ 8 & 24.8 $\pm$ 5.5 \\ 
\ion{Fe}{xix} & 13.50 $\pm$ 0.01 & 13.523$^{\mathrm{n}}$ & -500 $\pm$ 200 &   &   & 164 $\pm$ 9 & 55.3 $\pm$ 8.1 \\
\ion{Fe}{xx} & 12.815 $\pm$ 0.008 & 12.842$^{\mathrm{o}}$ & -600 $\pm$ 200 &   &   & 167 $\pm$ 8 & 56.7 $\pm$ 5.6 \\  
            \noalign{\smallskip}
            \hline
         \end{array}
     $$   
\begin{list}{}{}
\item[$^{\mathrm{a}}$] measured wavelength in $\rm \AA$ (in rest frame of NGC 3783)
\item[$^{\mathrm{b}}$] rest frame wavelength in $\rm \AA$
\item[$^{\mathrm{c}}$] blueshifted velocity in km s$^{\rm -1}$
\item[$^{\mathrm{d}}$] line flux in 10$^{\rm -5}$ photons cm$^{\rm -2}$ s$^{\rm -1}$
\item[$^{\mathrm{e}}$] line flux in Chandra spectrum as quoted in Kaspi et al. (\cite{kaspi2002}) in 10$^{\rm -5}$ photons cm$^{\rm -2}$ s$^{\rm -1}$
\item[$^{\mathrm{f}}$] measured equivalent width in m$\rm \AA$
\item[$^{\mathrm{g}}$] equivalent width in Chandra spectrum as quoted in Kaspi et al. (\cite{kaspi2002}) in m$\rm \AA$
\item[$^{\mathrm{h}}$] this value is yet to be benchmarked in the laboratory, and could be off by as much as 40 m$\rm \AA$, so the blueshift of the line could probably be about 500 km s$^{\rm -1}$ lower
\item[$^{\mathrm{i}}$] blended with a weaker absorption line of \ion{O}{v}
\item[$^{\mathrm{j}}$] partially filled in by re-emission
\item[$^{\mathrm{k}}$] perhaps partially filled in by re-emission
\item[$^{\mathrm{l}}$] blended with a weaker absorption line of \ion{Si}{xi}
\item[$^{\mathrm{m}}$] a blend of lines at 14.1548, 14.2042, 14.2120 and 14.2600~$\rm \AA$
\item[$^{\mathrm{n}}$] a blend of lines at 13.5070, 13.5146, 13.5210, 13.5540 and 13.5706~$\rm \AA$
\item[$^{\mathrm{o}}$] a blend of lines at 12.8130, 12.8270, 12.8319, 12.8470 and 12.9040~$\rm \AA$
\end{list}
  \end{table*}

 We observe an \ion{O}{vii} forbidden line in emission, but the signal-to-noise is not good enough to distinguish the accompanying intercombination and resonance lines. The fact that the forbidden line is more prominent than the other two indicates that they are formed in tenuous and photoionised gas (Porquet \& Dubau \cite{porquet}). The blueshift of the forbidden line, -600 $\pm$ 200 km s$^{\rm -1}$, is consistent with that of the warm absorber, which could support the idea that these lines originate from it. Kaspi et al. (\cite{kaspi2002}), on the other hand, find that this line is not blueshifted in their spectrum.

From investigating the narrow absorption lines with \emph{slab}, it became
 clear that we are dealing with two main ionisation ranges in the warm
 absorber. Kaspi et al. (\cite{kaspi2001}) also found multiple phases in the warm absorber, although the two phases that they model both correspond to our high-ionisation range (see Table~\ref{modelcomp}). Our high-ionisation phase gives rise to L-shell iron transitions, \ion{O}{viii}, \ion{O}{vii}, \ion{N}{vii} and \ion{C}{vi} absorption, whilst the low-ionisation gas is associated
 with M-shell iron forming an Unresolved Transition
 Array (UTA) at about 16.5~$\rm \AA$ (0.751~keV), as well as \ion{O}{iv}, \ion{O}{v} and \ion{O}{vi} lines. Both phases are blueshifted by about -800 $\pm$ 200 km s$^{\rm -1}$, a figure obtained by fitting the blueshift of each phase as a whole. This is consistent with the value of Kaspi et al. (\cite{kaspi2002}) (-590 $\pm$ 150 km s$^{\rm -1}$).

The many ions of iron apparent in the spectrum provide a tight constraint, independent of abundance, on the ionisation structure of the absorbing gas. Iron is used for this purpose as, in plasma at a given ionisation parameter, it will be present in several different ionisation states. Measuring the relative importances of these states in the spectrum gives a good measure of the ionisation parameter of the absorber. In general, as the plasma becomes less ionised, the main locus of iron absorption in the spectrum shifts to longer wavelengths.

From our identification of which iron states are present using \emph{slab}, there seem to be two locations where iron absorption is particularly important in this spectrum. The first of these is deep absorption from L-shell iron at around 13$-$15~$\rm \AA$ (0.954$-$0.827~keV), where we see \ion{Fe}{xvii}, \ion{Fe}{xviii}, \ion{Fe}{xix} and \ion{Fe}{xx}. At lower wavelengths, it becomes increasingly difficult to detect discrete absorption features from the more highly ionised states of iron.

If we consider the form of the spectrum around about 15~$\rm \AA$, where we might expect absorption from \ion{Fe}{xvi} and \ion{Fe}{xv}, we can see that there must be a far lower column of these ions as they would absorb away the low-wavelength side of the 15~$\rm \AA$ feature otherwise. We claim then that this represents the long-wavelength boundary of iron absorption from the high-ionisation phase of the warm absorber. 

The other important locus of iron absorption is the Unresolved Transition Array of M-shell iron at around 16$-$17~$\rm \AA$ (0.775$-$0.729~keV). This is the main spectral signature of the cold phase of the warm absorber. Once the blueshift of this phase has been established using the oxygen absorption lines which also originate from it, the position of the UTA in the spectrum can be used to determine the ionisation state of the gas. The high wavelength side of the 15~$\rm \AA$ feature then forms a boundary to the iron absorption from the low-ionisation phase.

We claim, then, that the 15~$\rm \AA$ feature provides us with a very clear cut-off between the iron absorption from two separate phases of gas. The height and shape of this feature places strict limits on exactly which iron states can be present in the absorbing medium.

Having established the existence of the two main ionisation regimes, we constructed a simple two phase warm absorber model using \emph{xabs} in order to gain some insight into the overall physical structure and composition of the absorbing gas. We also constructed a detailed model using \emph{slab} in order to obtain more precise measurements of the columns of individual ions within the absorber. These two models are described in the following sections.

\subsection{Two-phase warm absorber model}

The blueshift of the two phases was set to 800 km s$^{\rm -1}$. The narrow absorption lines are at the width of the RGS spectral resolution, so we cannot directly derive a turbulent velocity from them. For the purpose of these fits, we use a velocity width of 300 km s$^{\rm -1}$ (where the velocity width is the rms width, i.e. gaussian sigma, for each individual velocity component of each line), similar to that used in the analysis of the Chandra HETGS spectrum of this source (Kaspi et al. \cite{kaspi2001}).

The first stage was to determine the ionisation parameters of the two phases using \emph{xabs}. In the high-ionisation phase, if we take the most important iron states to be \ion{Fe}{xvii}, \ion{Fe}{xviii} and \ion{Fe}{xix}, we can estimate an overall log $\xi$ to be 2.4. In reality, of course, this is a simplification as the presence of the many higher iron states implies a range of ionisation parameters, from 2.4 upwards, in the high-ionisation phase. The lower-ionisation gas, on the other hand, is fairly well explained by gas at a single ionisation parameter. This is obtained from the position and form of the UTA, which is accurately reproduced with a log $\xi$ of 0.3.

Once we had the ionisation parameters of the two phases, we were able to fit overall equivalent hydrogen columns, assuming Anders \& Grevesse (\cite{anders}) elemental abundances. The abundances of individual elements were then allowed to vary. \emph{Xabs} includes the elements He, C, N, O, Ne, Na, Mg, Si, S, Ar and Fe; if it was not possible to establish the presence or abundance of a particular element, it was kept in the model but fixed at its solar abundance.

\begin{figure*}
   \centering
   \includegraphics[width=14cm]{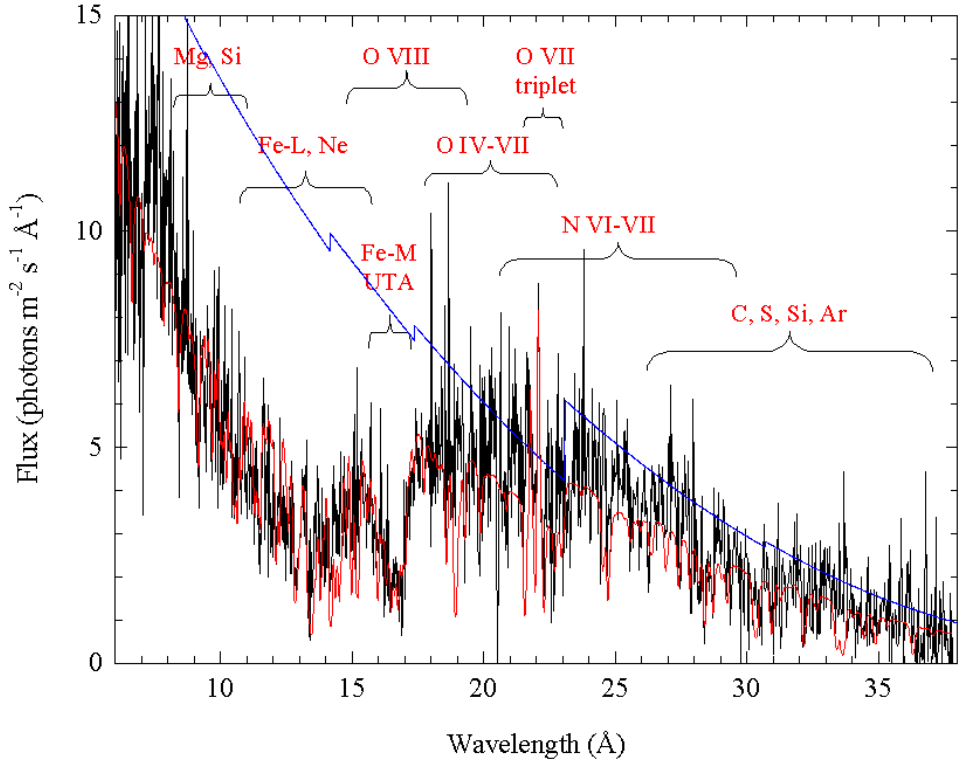}
      \caption{Combined RGS1 and RGS2 rest frame spectrum of NGC 3783 (black), with the power-law continuum (blue) and the power-law plus two-phase warm absorber model (red) superimposed, plotted over the range
      6$-$38~$\rm \AA$ (0.326$-$2.066~keV). Species absorbing or emitting in various spectral
      ranges are indicated.
              }
         \label{rgs_spec}
   \end{figure*}

\begin{figure*}
   \centering
   \includegraphics[width=14cm]{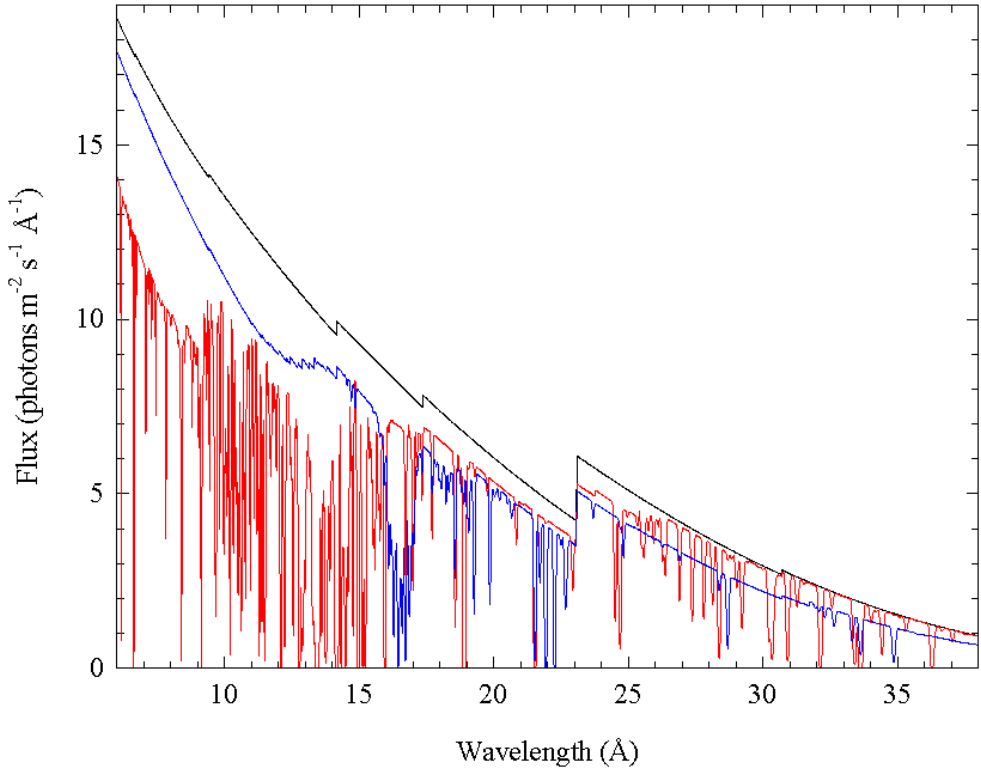}
      \caption{The two phases of the warm absorber model applied separately to the continuum (high-ionisation phase in red, low-ionisation phase in blue, and power-law continuum in black) plotted over the range
      6$-$38~$\rm \AA$ (0.326$-$2.066~keV; plotted in the rest frame). 
              }
         \label{model_comps}
   \end{figure*}

This simple two-phase model, superimposed on the data, is shown in Fig.~\ref{rgs_spec}, and Fig.~\ref{model_comps} shows the two phases of the absorber applied separately to the continuum. Table~\ref{hot} gives the values of the parameters derived for the two phases. It can be seen that the overall model reproduces the global form of the spectrum very well although there are discrepancies. The assumption of a single ionisation parameter for the entire high-ionisation phase of the gas leads to an overestimation of the iron absorption at the low-wavelength side of the 15~$\rm \AA$ feature and an underestimation of the iron absorption at wavelengths below this.

The ratios of the elemental abundances to each other in the high-ionisation phase are probably not too different from solar. Nitrogen appears to be underabundant, which is the opposite of what was found in IRAS 13349+2438 (Sako et al. \cite{sako}) and NGC 1068 (Kinkhabwala et al. \cite{kinkhabwala}). The solar oxygen abundance used in SPEX (Anders and Grevesse \cite{anders}) has been shown by a new measurement (Allende Prieto et al. \cite{allende}) to be too high, and thus our oxygen abundances are probably underestimated by about 0.2. In the low-ionisation phase, iron needs to be about ten times more abundant relative to C, N and O in order to reproduce the form of the very deep UTA. 

  \begin{table}
    
      \caption[]{Properties of the two-phase warm absorber model}
         \label{hot}
     $$
         \begin{array}{p{1.0in}p{1.0in}p{1.0in}}
            \hline
            \noalign{\smallskip}
            Parameter & High-ionisation phase & Low-ionisation phase \\
            \noalign{\smallskip}
            \hline
            \noalign{\smallskip}
            log $\xi$$^{\mathrm{a}}$  & 2.4 & 0.3 \\
            N$_{\rm H}$$^{\mathrm{b}}$  & 2.8 $^{\rm +0.01}_{\rm -0.3}$ x 10$^{\rm 22}$ & 5.4 $^{\rm +0.05}_{\rm -0.5}$ x 10$^{\rm 20}$ \\
            v$_{\rm turb}$$^{\mathrm{c}}$ & 300 & 300 \\
            v$_{\rm blue}$$^{\mathrm{d}}$ & 800 & 800 \\
            C abundance$^{\mathrm{e}}$ & 0.8 & 1.0 \\
            N abundance & 0.2 & 1.0 \\
O abundance & 0.7 & 1.0 \\
Ne abundance & 0.9 & \\
Mg abundance & 0.9 & \\
Si abundance & 2.5 & \\
S abundance & 1.0 & \\
Ar abundance & 0.7 & \\
Fe abundance & 0.9 & 10.7 \\

            \noalign{\smallskip}
            \hline
         \end{array}
     $$
\begin{list}{}{}
\item[$^{\mathrm{a}}$] Log of the ionisation parameter, where $\xi$ is in erg cm s$^{\rm -1}$
\item[$^{\mathrm{b}}$] Equivalent hydrogen column of phase in cm$^{\rm -2}$
\item[$^{\mathrm{c}}$] Turbulent velocity in km s$^{\rm -1}$
\item[$^{\mathrm{d}}$] Blueshift of phase in km s$^{\rm -1}$
\item[$^{\mathrm{e}}$] Abundances relative to solar; estimated errors on abundances are $\pm$ about 0.5. Values are only given for elements whose abundances could be directly measured; the abundance of all other elements in the model was fixed at 1
\end{list}{}{}
  \end{table}

%

%

\subsection{Detailed modelling}

\begin{figure*}
   \centering
   \includegraphics[width=18cm]{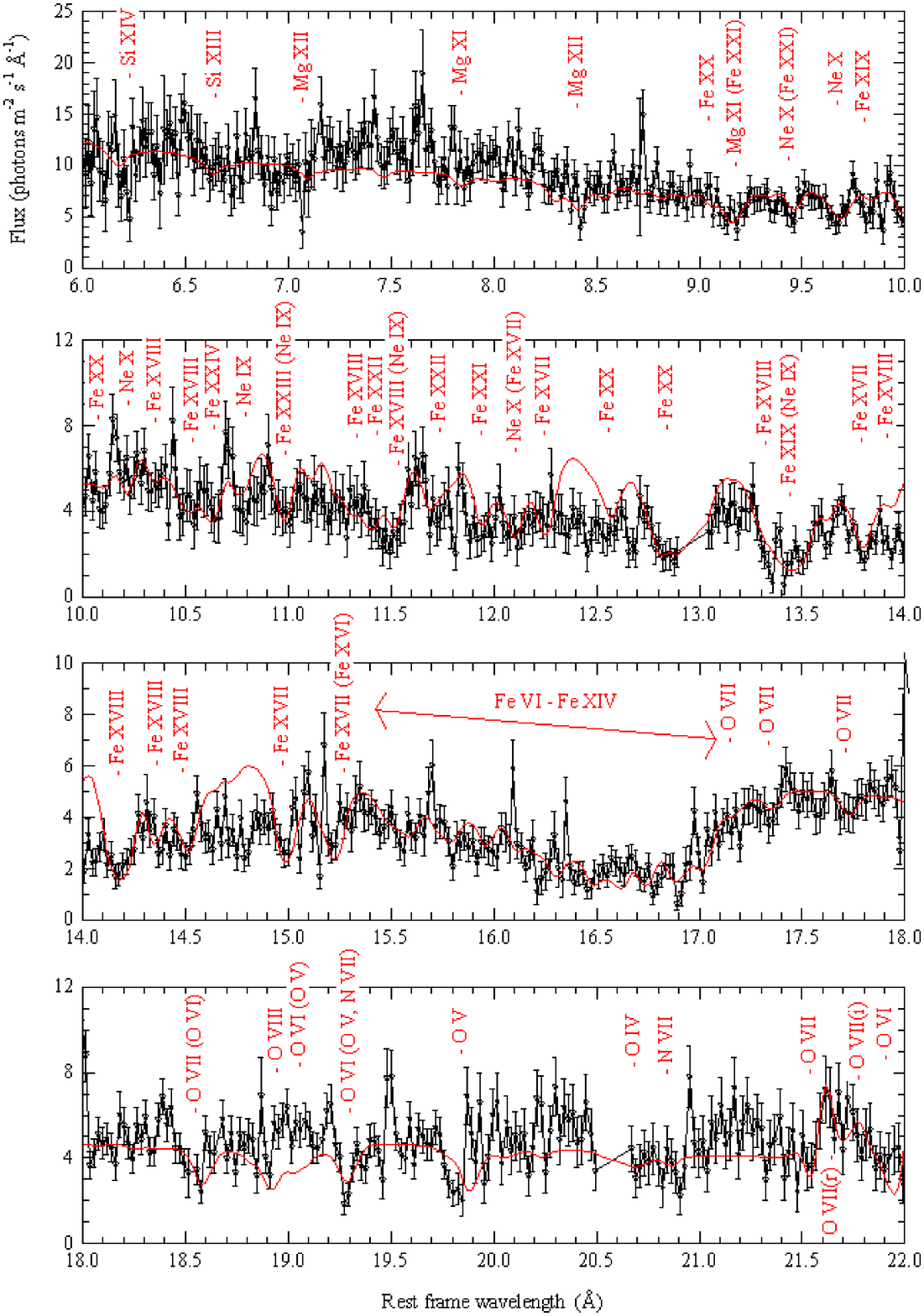}
      \caption{Combined RGS1 and RGS2 rest frame spectrum of NGC 3783 (black), binned
      into groups of two channels, with detailed warm absorber model superimposed (red), plotted over the range
      6$-$22~$\rm \AA$ (0.326$-$0.564~keV). The most important spectral features in the data and
      model are labelled.
              }
         \label{spec_1}
   \end{figure*}

\begin{figure*}
   \centering
   \includegraphics[width=18cm]{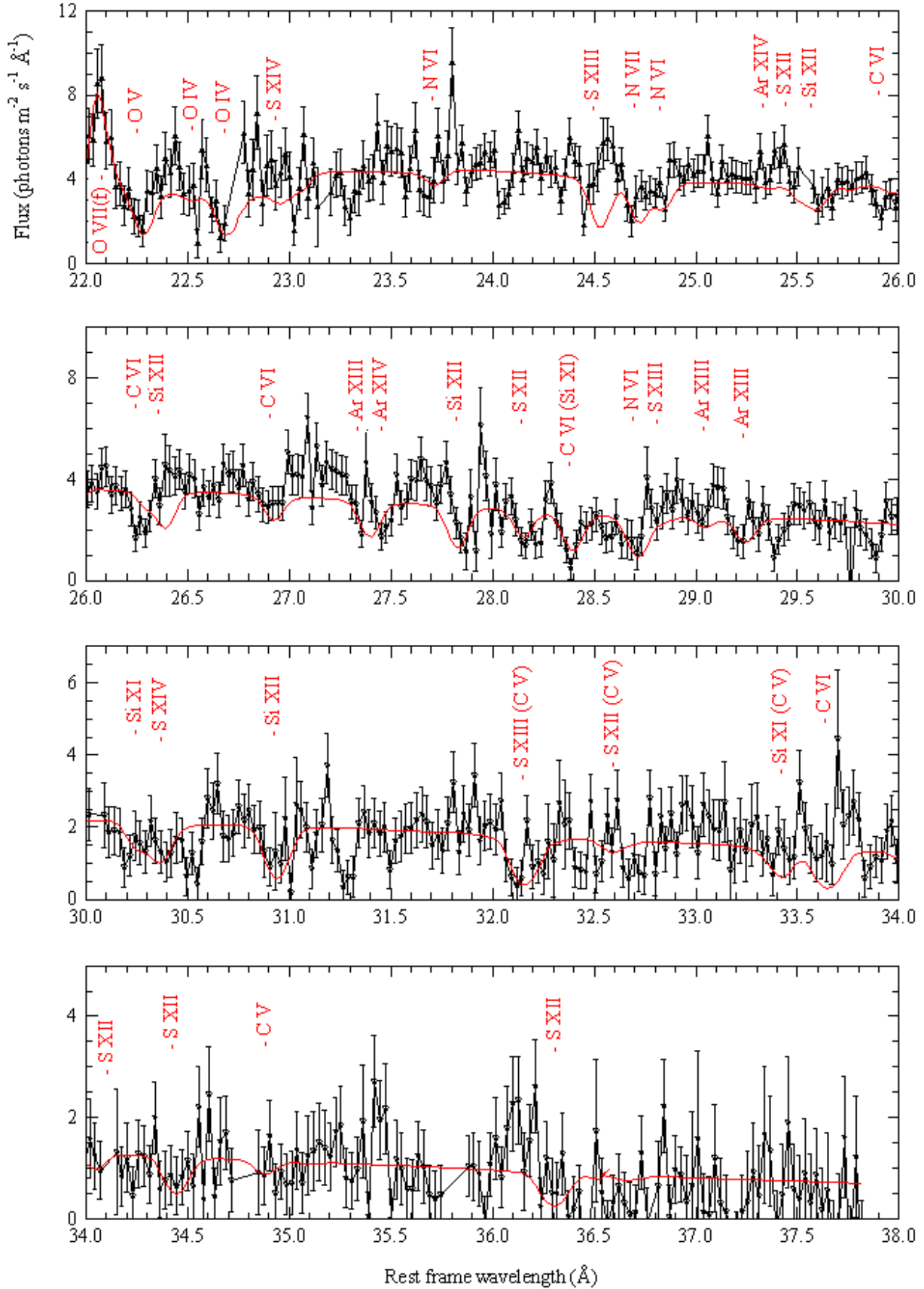}
      \caption{Combined RGS1 and RGS2 rest frame spectrum of NGC 3783 (black), binned
      into groups of two channels, with detailed warm absorber model superimposed (red), plotted over the range
      22$-$38~$\rm \AA$ (0.564$-$2.066~keV). The most important spectral features in the data and
      model are labelled.
                   }
         \label{spec_2}
   \end{figure*}

\begin{figure*}
   \centering
   \includegraphics[width=14cm]{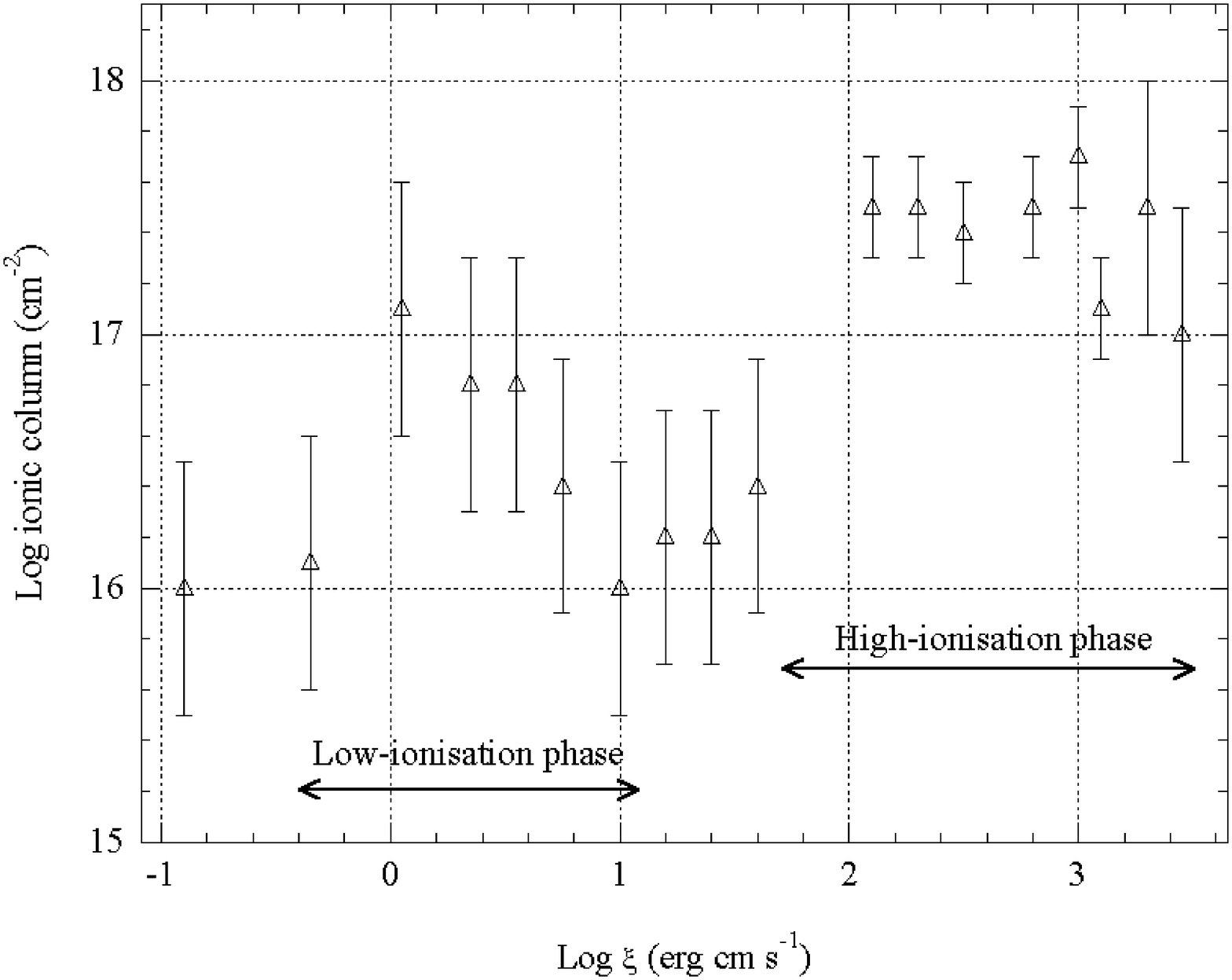}
      \caption{Ionic column plotted against ionisation parameter of maximum abundance for the iron states (except \ion{Fe}{xv}, for which only an upper limit could be obtained) listed in Table~\ref{ioncols}.  
              }
         \label{col_xi}
   \end{figure*}

The results of detailed modelling of the spectrum with \emph{slab} are shown in Fig.~\ref{spec_1} and Fig.~\ref{spec_2}. The individual ion columns and their estimated errors are listed in Table~\ref{ioncols}, alongside the log ionisation parameter at which the ion is most abundant. The L-shell iron absorption reproduces the form of the spectrum below 15~$\rm \AA$ very well in some regions, but there are still some significant residuals in others. Fig.~\ref{col_xi} shows the ionic column plotted against ionisation parameter of maximum abundance for the iron states (except \ion{Fe}{xv}, for which only an upper limit could be obtained) listed in Table~\ref{ioncols}. It clearly shows the existence of the two main ionisation ranges, and demonstrates that the low-ionisation gas is more tightly focussed at a certain level of ionisation than the high-ionisation phase, which extends over a range of ionisation parameters (note the logarithmic scale) - as already inferred from the discrepancies of the two-phase model.

\begin{table}
    
      \caption[]{The absorbing columns of individual ions in the detailed model of the warm absorber.}
         \label{ioncols}
     $$
         \begin{array}{p{0.8in}p{1in}p{0.6in}}
            \hline
            \noalign{\smallskip}
            Ion &  Log$_{\rm 10}$ column$^{\mathrm{a}}$ & Log $\xi$$_{\rm ma}$$^{\mathrm{b}}$ \\
            \noalign{\smallskip}
            \hline
            \noalign{\smallskip}
\ion{C}{v} & 16.5 $\pm$ 0.5 & 0.15 \\
\ion{C}{vi} & 17.5 $\pm$ 0.2 & 1.15 \\
\ion{N}{vi} & 16.9 $\pm$ 0.2 & 0.70 \\
\ion{N}{vii} & 16.8 $\pm$ 0.2 & 1.50 \\
\ion{O}{iv} & 17.1 $\pm$ 0.2 & -0.75 \\
\ion{O}{v} & 17.5 $\pm$ 0.2 &  -0.05 \\
\ion{O}{vi} & 17.0 $\pm$ 0.5 & 0.45 \\
\ion{O}{vii} & 17.3 $\pm$ 0.2 & 1.15 \\
\ion{O}{viii} & 16.8 $\pm$ 0.5 & 1.75 \\
\ion{Ne}{ix} & 18.0 $\pm$ 0.5 & 1.70 \\
\ion{Ne}{x} & 18.2 $\pm$ 0.5 & 2.25 \\
\ion{Mg}{xi} & 17.5 $\pm$ 0.5 & 2.10 \\
\ion{Mg}{xii} & 17.8 $\pm$ 0.5 & 2.60 \\
\ion{Si}{xi} & 17.0 $\pm$ 0.2 & 1.70 \\
\ion{Si}{xii} & 17.9 $\pm$ 0.2 & 2.05 \\
\ion{Si}{xiii} & 17.5 $\pm$ 0.2 & 2.40 \\
\ion{Si}{xiv} & 18.0 $\pm$ 0.5 & 2.90 \\
\ion{S}{xii} & 17.0 $\pm$ 0.2 & 1.70 \\
\ion{S}{xiii} & 17.5 $\pm$ 0.2 & 2.05 \\
\ion{S}{xiv} & 16.7 $\pm$ 0.2 & 2.40 \\
\ion{Ar}{xiii} & 16.1 $\pm$ 0.2 & 1.80 \\
\ion{Ar}{xiv} & 16.4 $\pm$ 0.5 & 2.10 \\
\ion{Fe}{vi} & 16.0 $\pm$ 0.5 & -0.90 \\
\ion{Fe}{vii} & 16.1 $\pm$ 0.5 & -0.35 \\
\ion{Fe}{viii} & 17.1 $\pm$ 0.5 & 0.05 \\
\ion{Fe}{ix} & 16.8 $\pm$ 0.5 & 0.35 \\
\ion{Fe}{x} & 16.8 $\pm$ 0.5 & 0.55 \\
\ion{Fe}{xi} & 16.4 $\pm$ 0.5 & 0.75 \\
\ion{Fe}{xii} & 16.0 $\pm$ 0.5 & 1.00 \\
\ion{Fe}{xiii} & 16.2 $\pm$ 0.5 & 1.20 \\
\ion{Fe}{xiv} & 16.2 $\pm$ 0.5 & 1.40 \\
\ion{Fe}{xv} & 15.4$^{\mathrm{c}}$ & 1.60 \\
\ion{Fe}{xvi} & 16.4 $\pm$ 0.5 & 1.60 \\
\ion{Fe}{xvii} & 17.5 $\pm$ 0.2 & 2.10 \\
\ion{Fe}{xviii} & 17.5 $\pm$ 0.2 & 2.30 \\
\ion{Fe}{xix} & 17.4 $\pm$ 0.2 & 2.50 \\
\ion{Fe}{xx} & 17.5 $\pm$ 0.2 & 2.80 \\
\ion{Fe}{xxi} & 17.7 $\pm$ 0.2 & 3.00 \\
\ion{Fe}{xxii} & 17.1 $\pm$ 0.2 & 3.10 \\
\ion{Fe}{xxiii} & 17.5 $\pm$ 0.5 & 3.30 \\
\ion{Fe}{xxiv} & 17.0 $\pm$ 0.5 & 3.45 \\

            \noalign{\smallskip}
            \hline
         \end{array}
     $$
\begin{list}{}{}   
\item[$^{\mathrm{a}}$] where the column is in cm$^{\rm -2}$
\item[$^{\mathrm{b}}$] log $\xi$ at which the ion is at its maximum abundance ($\xi$ is in erg cm s$^{\rm -1}$)
\item[$^{\mathrm{c}}$] upper limit
\end{list}{}{}
  \end{table}

%
%
%

\section{Discussion and conclusions}

\subsection{Ultraviolet image}

The XMM OM provides us with the first ever UV image of the nuclear regions of NGC 3783. Although the active nucleus is by far the brightest source of ultraviolet light in the galaxy, 
 our image also shows UV emission from the spiral arms, which presumably traces the location of 
 star formation activity. Although 
 the spiral arms are visible in UV, the bar of the galaxy (oriented approximately North-to-South 
 in our image) is not 
 strongly present, indicating that star formation is much less 
 important in the bar than it is in the arms, or alternatively that dust is obscuring the UV emission 
 from the bar. 

\subsection{Lightcurves}

It is clear from Fig.~\ref{ltcv} that there is unlikely to be sufficient spectral variability in this source to have a serious effect on our modelling of the continuum underlying the RGS spectrum. However, there is clearly variability in the flux and spectral shape on many different timescales. Intensity variations on the scale of about one to three hours, which we see most prominently in the 0.2$-$2~keV band, seem to modulate both soft and hard bands in the same way. Also, the changes that we observe in the hardness ratio demonstrate variability in the relative intensities of the two bands over the course of a few hours. Fig.~\ref{colour} indicates that the spectrum generally softens as the source intensity increases, although at the highest intensity point, that is at the very end of our observation, the source is actually in a harder state. The UV lightcurve is almost completely flat, within the errors, implying less than 10\% variability in the UV flux from the nucleus during our observation. There is therefore no evidence in our dataset for a direct correlation between UV and X-ray variability over the course of a few hours.

\subsection{Continuum and iron K$\alpha$ line}

We find that the 0.2$-$10~keV spectral continuum of NGC 3783 is dominated by a power-law at high energies, whilst there is significant absorption by ionised gas - the warm absorber - at lower energies. There is no evidence for a soft excess, although of course the presence of a soft excess could be masked by the warm absorber. There is also no requirement for a reflection component in our spectrum.

The iron K${\alpha}$ emission line has two components; one narrow component at 6.40 keV, and a broader component which is redshifted by about 5000 km s$^{\rm -1}$. The profile of this composite line looks superficially similar to that of the NGC 3783 spectrum (1) of Nandra et al. (\cite{nandra}), although our broad component is far narrower than the one they observed. We were able to fit our broad component with a gaussian; it does not contain enough flux to obtain good constraints on a relativistic disc line model. Our narrow Fe K${\alpha}$ line is apparently similar to that observed by Kaspi et al. (\cite{kaspi2002}), although our derived parameters are not well constrained. 

The broad component of the Fe K${\alpha}$ line implies emission from gas that is infalling close to the nucleus. The line broadening would presumably be dynamic rather than thermal, as Fe K${\alpha}$ arises from fluorescence of neutral iron, so the emitting matter is in a region of high speed motion under strong gravity. Certainly, the derived velocity broadening of this component, at 11000 $\pm$ 4000 km s$^{\rm -1}$, would place it in the very broad line region of Winge et al. (\cite{winge}). We were unable to ascertain whether the broadening is actually relativistic, although if we were dealing with emission from close enough to the black hole to be relativistically broadened, we might expect the iron to be rather more highly ionised than the energy of this line implies.

That the broad component of the Fe K$\alpha$ line was not observed by Kaspi et al. (\cite{kaspi2002}) in the 900 ks Chandra observation, where the signal-to-noise in this spectral region was very good, may imply that the spectrum has changed between the two observations. Infalling gas close to the central engine of the AGN, but not yet close enough to become too hot to emit Fe K${\alpha}$, might be a transitory phenomenon which is not always observable.

The Fe K${\alpha}$ emission is accompanied by another line at higher energy, which is probably a combination of Fe K${\beta}$ and \ion{Fe}{xxvi}. The apparent broadening and redshift of this line are, then, a function of the flux and redshift of the components that it contains. \ion{Fe}{xxvi} emission would not originate from the same material as Fe K${\alpha}$; instead, this must be emission from much hotter gas. 

\subsection{Warm absorber}

\subsubsection{Modelling strategy}

We developed a strategy for modelling the complex RGS spectrum of this object using SPEX, with the aim of reproducing both line and continuum absorption self-consistently, and of obtaining some physical insight into the properties of the warm absorber.

This process has several steps. Firstly, \emph{slab} - a model which applies absorption by individual ions in a photoionised medium to a given spectral continuum - is used to identify the patterns of line absorption by different ionic species in the warm absorber, and to measure the blueshift of the lines. This model is also used to investigate the contributions that each ion can make to continuum absorption. Once the lines have been identified, the relative strengths of line absorption by the different iron species are used to derive the ionisation structure of the absorber. Iron is used for this purpose due to its wide range of ionisation states, several of which can be observable in gas at a given level of ionisation, thus providing a good constraint on the ionisation parameter of the medium. We contend that this gives a more reliable picture of the ionisation structure of the absorber than a simple measurement of \ion{O}{vii} and \ion{O}{viii} edge opacities, which is dependent on continuum absorption (a more ambiguous diagnostic than line absorption) from only two ions. 

At this stage, the \emph{xabs} model, which applies line and continuum absorption by a given column of photionised gas at a given ionisation parameter to the spectral continuum, is used to fit an overall column for the absorber at the ionisation parameter independently derived from the iron absorption. Multiple \emph{xabs} models can be used to represent the different ionisation phases within the absorber. This is done with the assumption of solar abundances within the absorber, but, once an estimate of the overall column has been derived, the individual elemental abundances can be allowed to vary. To get columns for the individual ionic states, one can return to \emph{slab} to fit values for these. 

At every stage of the process outlined here, due to the likelihood of \lq blind\rq \, $\chi$$^{\rm 2}$ fitting being misled by the enormous complexity of the spectrum and model, it is necessary to spend some time getting to know the model properties by trial and error. A probable range of realistic values needs to be established before using $\chi$$^{\rm 2}$ minimisation to home in on the final value for a parameter, being careful to ignore any spectral regions that will bias the fit towards unphysical results.

\subsubsection{Two-phase warm absorber model}

The two-phase warm absorber model reproduces the global form of the spectrum very well. It indicates that the ratios of elemental abundances to each other in the high-ionisation phase of the gas are probably not too different from solar. However, in order to fit the very deep UTA associated with the low-ionisation phase of the gas, iron would be ten times more abundant, relative to its solar value, than oxygen. This is sensitive to the turbulent velocity and overall absorbing column assumed for the low-ionisation phase, and the level of the spectral continuum itself. If the spectral continuum was higher (especially at the long wavelength end of the spectrum, where continuum absorption by \ion{C}{iv} and \ion{C}{v} becomes important), the equivalent hydrogen column of the low-ionisation phase would need to be higher and thus the iron abundance could be a lot lower. The value of $\xi$ in the model is too high for some of the lower charge states of Fe, therefore requiring high Fe abundance to compensate for the low fractional ionic abundances that the model produces. The low-ionisation phase might also have a wider distribution of $\xi$. Moreover, the ionisation balance calculations of Fe-M are very uncertain; in particular, the dielectronic recombination rates for these ions have never been measured or calculated with modern codes. 

The much shallower UTA in IRAS 13349+2438 (Sako et al. \cite{sako}) requires an iron abundance 2$-$3 times that of oxygen, and the UTA in the LETGS spectrum of NGC 5548 (Kaastra et al. \cite{kaastra2002b}) also suggests an overabundance of iron. It is possible that these apparent high iron abundances could be due to a covering factor effect (e.g. Arav et al. \cite{arav}). The iron abundance in the UTA is effectively measured against the abundances of C, N and O, and if these lines are highly saturated their equivalent widths will be determined by the covering factor and velocity width, and will be only weakly dependent on the column density.

The position of the UTA also coincides with the \ion{O}{vii} absorption edge (16.78~$\rm \AA$ (0.7389~keV) in the rest frame), and the deeper this edge, the lower the iron abundance required to fit the UTA. We would not expect significant \ion{O}{vii} absorption to originate from the low-ionisation phase itself, as it is not highly ionised enough. The high-ionisation phase, on the other hand, could be a plausible source of \ion{O}{vii} edge absorption. At the derived ionisation parameter of this phase, though, oxygen will be predominantly in the form of \ion{O}{viii}. Certainly, the fitted equivalent hydrogen column of the high-ionisation phase, at 2.8 x 10$^{\rm 22}$ cm$^{\rm -2}$, is easily high enough to give rise to significant edge absorption. Examination of the detailed form of the spectrum where the edges are expected, however, shows that any such edges would be completely masked by L-shell and M-shell iron absorption. We propose, then, that although \ion{O}{vii} and \ion{O}{viii} edge opacity is certainly part of the picture, absorption by L-shell and M-shell iron is likely to be more important in determining the detailed form of the spectrum in these regions.

The \ion{O}{viii} absorption is, in fact, seriously overpredicted in the two-phase model (N$_{\rm \ion{O}{viii}}$ $\geq$ 10$^{\rm 18}$ cm$^{\rm -2}$), implying a very deep \ion{O}{viii} Ly$\alpha$ absorption line (at 18.97~$\rm \AA$ (0.6536~keV), see Fig.~\ref{rgs_spec}) that is not matched by the data. It is of course possible that this line is partially filled in by \ion{O}{viii} line emission, but we cannot test this with the current dataset. It can also be seen from Fig.~\ref{rgs_spec} that the iron absorption is overpredicted at the low-wavelength side of the 15~$\rm \AA$ feature and increasingly underpredicted at lower wavelengths. Fig.~\ref{model_comps} shows that the two-phase model compensates for this by generating neon edge absorption at $\sim$ 9.5~$\rm \AA$. This clearly demonstrates that it was not correct to approximate log $\xi$ of the high-ionisation phase with a single value of 2.4 - there is a lot of gas which is more highly ionised than this, so the high-ionisation phase should be more accurately modelled as containing a range of ionisation parameters from 2.4 upwards. The low-ionisation phase, too, is not perfectly explained by a single phase with log $\xi$ of 0.3, but the single ionisation parameter is a good approximation in this case. 

A comparison between the parameters of our two-phase model, and that of Kaspi et al. (\cite{kaspi2001}), is given in Table~\ref{modelcomp}. Their low-ionisation phase corresponds most closely to our high-ionisation phase, whilst their high-ionisation phase is much more highly ionised than ours. It probably corresponds to the highly-ionised iron at the low-wavelength end of our spectrum which our two-phase model does not explain. Kaspi et al. (\cite{kaspi2001}) do not model our low-ionisation (UTA) phase, although they report the presence of the UTA. The warm absorber modelled by De Rosa et al. (\cite{derosa}) is consistent with the low-ionisation component of Kaspi et al. (\cite{kaspi2001}), and thus with our high-ionisation component.

The two models, although different, are not in fundamental disagreement with each other. The differences between them may be due to the methods used to model the spectrum. The Kaspi et al. (\cite{kaspi2001}) modelling relies on the use of measured equivalent widths of the absorption lines as well as the fitting of a continuum model to \lq line-free zones\rq \, where absorption lines are not expected. We, on the other hand, use the line absorption as the prime diagnostic and tie the continuum absorption directly to this. The elemental abundances in their model were kept at solar values, whilst ours were allowed to vary. These two phase models, though, are just approximations to a multi-phase absorber, as we discuss above, and as already pointed out in Behar \& Netzer (\cite{behar2002}).

\begin{table}
    
      \caption[]{Comparison of the properties of the two-phase warm absorber model presented here with that of Kaspi et al. (\cite{kaspi2001})}
         \label{modelcomp}
     $$
         \begin{array}{p{0.9in}p{0.5in}p{0.9in}p{0.7in}}
            \hline
            \noalign{\smallskip}
            Phase & Property & Our model & Kaspi et al. \cite{kaspi2001} model \\
            \noalign{\smallskip}
            \hline
            \noalign{\smallskip}
            High-ionisation & log $\xi$$^{\mathrm{a}}$  & 2.4 & 3.5 \\
            High-ionisation & N$_{\rm H}$$^{\mathrm{b}}$  & 2.8 $^{\rm +0.01}_{\rm -0.3}$ x 10$^{\rm 22}$ & 1.6 x 10$^{\rm 22}$ \\
            Low-ionisation & log $\xi$$^{\mathrm{a}}$  & 0.3 & 2.5 \\
            Low-ionisation & N$_{\rm H}$$^{\mathrm{b}}$  & 5.4 $^{\rm +0.05}_{\rm -0.5}$ x 10$^{\rm 20}$ & 1.6 x 10$^{\rm 22}$ \\
            \noalign{\smallskip}
            \hline
         \end{array}
     $$
\begin{list}{}{}
\item[$^{\mathrm{a}}$] Log of the ionisation parameter, where $\xi$ is in erg cm s$^{\rm -1}$
\item[$^{\mathrm{b}}$] Equivalent hydrogen column of phase in cm$^{\rm -2}$
\end{list}{}{}
  \end{table}

\subsubsection{Detailed modelling}

Our detailed model, using \emph{slab}, provides a list of the total columns of the main ions present in the multi-phase warm absorber, summed across all phases. This model (Fig.~\ref{spec_1} and Fig.~\ref{spec_2}) provides a better reproduction of the detailed form of the spectrum than the simple \emph{xabs} two-phase approximation. The agreement of our model with the data in the region of L-shell iron absorption, in particular, is excellent in some ranges. There are still some significant residuals in this region, though. The wavelengths and strengths of the L-shell iron absorption lines and blends are very well known, so it is unlikely that this is the source of the discrepancy. Some of the absorption may originate from elements, particularly calcium, which are not yet included in our model. It is possible to reduce some of the residuals by increasing the \ion{O}{viii} continuum absorption, although this then introduces very deep \ion{O}{viii} absorption lines which do not appear in our data. One interesting explanation for these residuals might be that our model lacks the higher level transitions associated with the UTA (Behar et al. \cite{behar}) which appear when there is a high column density in the UTA ions.

The derived columns of the L-shell iron states above \ion{Fe}{xx} do indicate that there is a lot of gas in the hot phase of this warm absorber at log ionisation parameters above 2.4, as shown in Fig.~\ref{col_xi}. This may be part of the explanation for the inability of the two-zone model to accurately reproduce the \ion{O}{viii} absorption; much of the gas in the hot phase may simply be too highly ionised for \ion{O}{viii} to form in significant quantities. The derived columns of \ion{O}{vii} and \ion{O}{viii} ($\sim$ 10$^{\rm 17}$ cm$^{\rm -2}$) in this detailed model are too low to produce significant edge absorption, but their values are fairly uncertain.

The effective area of the HETGS system falls off above around 23~${\rm \AA}$ (0.539~keV),
 so since the RGS spectrum extends up to ~38 ${\rm \AA}$ (0.326~keV), we can add quantitively to the knowledge gained from the Chandra spectrum of NGC 3783. In this high wavelength range we see absorption lines from \ion{N}{vi}, \ion{N}{vii}, \ion{C}{vi} and possibly \ion{C}{v}, \ion{Si}{xi}, \ion{Si}{xii}, \ion{S}{xii}, \ion{S}{xiii}, \ion{S}{xiv}, \ion{Ar}{xiii} and \ion{Ar}{xiv}. Direct superimposition of the model on the spectrum (Fig.~\ref{spec_1} and Fig.~\ref{spec_2}) reveals clearly the inaccuracies in the model wavelengths of various lines, and shows that there are further absorption lines in the spectrum that are yet to be included in our model. This work will be carried forward using the 280 ks RGS exposure of NGC 3783.

What is the connection between the higher and lower ionisation phases
 of the warm absorber? They appear to be flowing towards us at a similar
 speed, around 800 km s$^{\rm -1}$. The high-ionisation phase has an equivalent hydrogen
 column about fifty times that of the low-ionisation phase in our line of sight,
 and the low-ionisation phase apparently has a very high iron abundance. The spread of values of the ionisation parameter in the two phases could give us a clue as to what is going on. As indicated by the results of the \emph{xabs} modelling, and also in Fig.~\ref{col_xi}, the low-ionisation phase appears to be concentrated around a single ionisation parameter, whilst the high-ionisation phase contains a range of levels of ionisation. When the central engine of NGC 3783 is radiating at a given luminosity, the ionisation parameter of the absorbing outflow is determined by the density and distance from the source of the absorbing gas. $\xi$ is more sensitive to changes in distance than changes in density. One could, then, surmise that the wide range of $\xi$ in the high-ionisation phase implies absorption by gas at a wide range of distances and (to a lesser extent perhaps) densities, whereas the low-ionisation, UTA phase absorption is quite localised to gas at a certain distance from the source, and at a particular density. This would imply the existence of a highly-ionised, multi-temperature outflow (as in the multi-temperature wind model posited by Krolik \& Kriss \cite{krolik}), spread over a range of distances from the source, and denser and perhaps iron-rich material at a fixed location within the circumnuclear environment. 

Sako et al.
 (\cite{sako}) claim that the low-ionisation phase of the warm absorber in IRAS 13349+2438 might
 originate in the dusty torus surrounding the active nucleus, which would fit in with the scenario described here. Perhaps
 the abnormally high iron content of the corresponding phase in NGC 3783 is
 due to the sublimation of iron-rich dust from the edge of a dusty
 torus. If this was the case, the high-ionisation phase would form the bulk of
 the warm absorber, and wherever it encountered the edge of the dusty
 torus it would sweep up denser material; this would then be
 photoionised and the dust it contained would be sublimated, giving
 rise to the high iron content of the low-ionisation phase. This places
 constraints, of course, on the composition of the dust, which
 would have to be justified. The higher signal-to-noise RGS spectrum which has now been obtained will allow us to investigate the
 relative abundance of other non-volatile elements which can be
 bound into dust grains (including silicon and carbon), to test
 this hypothesis further.


\begin{acknowledgements}
      This work is based on observations obtained with XMM-Newton, an ESA
science mission with instruments and contributions directly funded by ESA
Member States and the USA (NASA). The MSSL authors acknowledge the support
of PPARC. SRON is supported financially by NWO, the
Netherlands Organization for Scientific Research. Thanks are due to F. Haberl for providing a PN small window mode response matrix. 
\end{acknowledgements}

\end{document}